# Rapid direct growth of graphene on single-crystalline diamond using nickel as catalyst

N. Suntornwipat[1*], A. Aitkulova[1], V. Djurberg[1] and S. Majdi[1]

[1] *Division for Electricity, Department of Electrical Engineering, Uppsala University, Box 65, SE-751 03, Uppsala, Sweden.*

*Corresponding author: Nattakarn.Suntornwipat@angstrom.uu.se

**Abstract**

Although theoretical investigations indicate that the successful combination of graphene and diamond would give interesting properties, only a limited number of reports dealing with the subject have been published. Here, we present a rapid thermal process (RTP) which involves Ni as metal catalyst for a direct growth of graphene on diamond at a temperature of 800 °C for 1 min. This process operates with a combination of a lower temperature and for a shorter duration than what has previously been reported. Thin Ni films with different thicknesses were deposited on top of (100) single-crystalline diamond. After RTP, the coverage of monolayer graphene was found to be around 20% shown by the intensity ratio between the 2D- and G-peak using Raman spectroscopy on 50-nm thick Ni films. In addition, x-ray photoelectron spectroscopy and atomic force microscopy analysis were conducted for additional investigation. For electrical characterization, Hall-effect measurements were performed at temperatures between 80 and 360 K.

## Introduction

Graphene, with only a single atomic layer of carbon and diamond, the three-dimensional hardest carbon allotrope both reveal extreme electronic properties. Over the last decade there has been a growing interest in combining these two to create carbon-carbon hybrids. Diamond is a wide bandgap semiconductor with excellent material properties. High thermal conductivity (24 W/cmK) [1] and high carrier mobilities (4500 and 3800 $cm^2$/Vs for electrons and holes respectively) [2] at room temperature (RT) as well as a high electric breakdown field (9.5 MV/cm) [3] and low dielectric constant (5.7) are just a few of the properties highlighting the material. For that reason, diamond has been extensively studied for use in applications related to power electronics [4], spintronics [5], nitrogen vacancy centers [6] and valleytronics [7,8]. The reports do not only suggest diamond as an excellent substrate for graphene [9] but also that the current-carrying capability of graphene-based devices can be substantially higher with substrates made out of diamond compared to $SiO_2$/Si [10]. This is attributed to diamond's high thermal conductivity [10]. In addition, graphene has many desirable semiconducting properties [10–14] and it has been proposed that the combination of graphene and diamond could enable bandgap separation in graphene [9,15], inject electronic spin [15], and be used for spin-polarized conducting wires [16].

Even though there are indications that graphene-diamond heterostructures would be very beneficial in many applications only a limited number of reports on the fabrication of graphene layers on diamond exist. To avoid the complexity of graphene transfer it is desirable to grow the graphene directly on the diamond substrate. This has been attempted on single-crystalline (SC) diamond with a metal catalyst (Cu [17], Ni [18,19] and Fe [20]) and without a catalyst [17–19,21–23] . The use of a catalyst has proven to accelerate graphene growth [24]. By heating in vacuum at 950 °C for 90 min a graphene coverage of 85% has been obtained on a (100) SC diamond sample using Cu as the catalyst [17]. Hall-effect measurements with a van der Pauw configuration showed a sheet hole concentration and hole mobility at RT of $12 \times 10^{13}$ cm$^{-2}$ and 410 cm$^2$/Vs respectively [17]. Kanada et al. [19] reported a growth of mainly multilayer graphene (~75%) on (111) SC diamond using Ni heated at 900 °C for 1 min in Ar$_2$ atmosphere. At RT, they reported a sheet hole concentration and hole mobility of $5.7 \times 10^{13}$ cm$^{-2}$ and 140 cm$^2$/Vs respectively.

In this study, we present a fast process to grow graphene directly on (100) single-crystalline chemical vapor deposition (SC-CVD) diamond using Ni as catalyst at a significantly lower temperature, 800 °C, for the short time of 1 min. The thickness of the Ni film was varied from 50 to 300 nm. The graphene layer was analyzed with Raman spectroscopy, 3D coherence scanning interferometry, x-ray photoelectron spectroscopy (XPS) and atomic force microscopy (AFM). Moreover, Hall-effect measurements were performed to investigate the electrical properties.

**Experimental details**

Commercially available free-standing optical-grade SC-CVD diamond plates from Element six Ltd. were used as substrates. The samples (3.35 × 3.35 mm$^2$) were homoepitaxially synthesized in the (100) crystallographic orientation with a thickness of 500 µm and a surface roughness below 5 nm. Prior to the process, the samples were treated in a graphite etch (HNO$_3$:HClO$_4$:H$_2$SO$_4$) at 180 °C for 40 minutes followed by surface oxygen-termination in a mild oxygen plasma for 60 seconds. The Ni film was deposited directly on top of the diamond substrate using electron-beam evaporation with a base pressure below $2 \times 10^{-6}$ Torr. The film thickness was varied from 50 to 300 nm. Rapid thermal annealing was used for 1 min in a nitrogen (N$_2$) atmosphere to form the graphene. The annealing temperature was set to 800 °C as previous studies found the number of graphene layers to increase with temperature for temperatures between 800 and 1000 °C [21]. A mixture of H$_2$SO$_4$ and H$_2$O$_2$ removed the Ni film which resulted in only a graphene layer on top of diamond remaining. After, the sample surfaces were analyzed by applying 3D coherence scanning interferometry (ZYGO NexView), AFM in Peak Force Tapping mode (Bruker Dimension Icon ICON4-SYS), XPS (Physical Electronics Quantera II Scanning XPS Microprobe) and Raman spectroscopy (Renishaw InVia confocal Raman microscope) with a laser using 532 nm excitation wavelength.

Hall-effect measurements were performed at a temperature range of 80 to 360 K on a Hall-bar with a length and width of 700 and 200 µm, respectively. An AC Hall system consisting of two Signal Recovery 7265 DSP lock-in amplifiers and an electromagnet applying a magnetic field between −0.5 and 0.5 T perpendicular to the electric field were used. The samples were mounted in a sample holder in a Janis ST-300MS vacuum cryostat, and a calibrated GaAlAs diode in conjunction with a Lake

Shore 331 temperature controller were used for temperature control using liquid nitrogen as the coolant.

## Results and discussion

After heating the sample and removing the Ni film, Raman spectroscopy and XPS were used to verify the graphene growth. A comparison using Raman spectroscopy before and after the process is shown in Fig. 1(a). We observe four main peaks at approximately 1334, 1422, 1584 and 2703 cm$^{-1}$ which are contributed to the first order zone-center vibrational mode of diamond [25,26], diamond-like carbon (DLC) species [17], G- and 2D-peak [27,28] respectively. The peaks at 1334 and 1422 cm$^{-1}$ are not associated with the graphene layer and are also observed on our diamond substrate. Raman observations (Fig. 1(a)) reveal that the intensity ratio between the 2D- and G-peak ($I_{2D}/I_G$) is approximately 3.3 and the full width at half maximum (FWHM) of the G- and 2D-peak is ~ 21 and 28 cm$^{-1}$, designating monolayer graphene [28,29].

XPS, with a quartz crystal monochromator, a beam size of 200 µm in diameter and a power of 50 W was used to analyze the carbon-bonding characteristics of our sample. A survey spectrum and a C1s core level spectrum were acquired with a pass energy of 224 and 55 eV, respectively. A dual beam charge neutralization system (a low-energy electron beam and a low-energy Ar$^+$ ion gun) was employed during XPS spectrum acquisition. We observe atomic concentrations of mainly carbon (C1s) and oxygen (O1s), whereas less than 5% silicon (Si2s, Si2p) and argon (Ar2p) are observed on the graphene-on-diamond sample, as seen in the inset to Fig. 1(b). The argon and oxygen contents are attributed to our charge neutralization system and due to the film exposure to the ambient conditions, respectively.

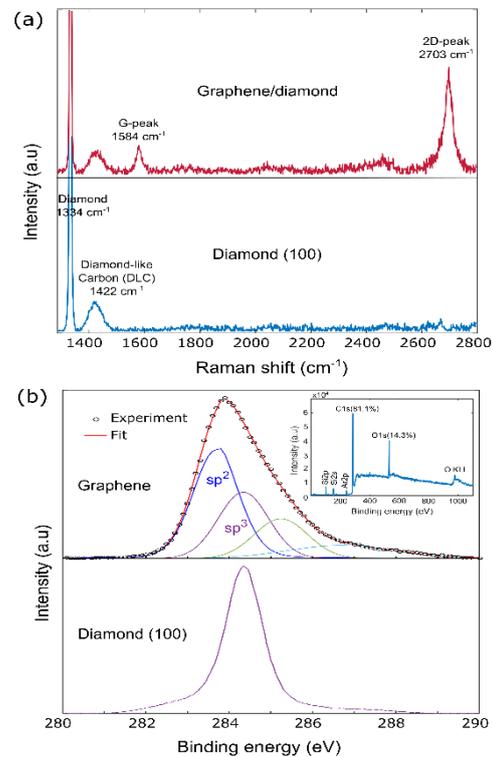

**Fig. 1.** (a) Raman spectroscopy on SC-CVD diamond with (100)-orientation compared with that of graphene formed on diamond using annealing of Ni with the thickness of 50 nm at 800 °C for 1 min. (b) XPS results of the C1s core level spectrum with a pass energy of 55 eV for both the graphene-on-diamond sample and the diamond substrate. The inset shows the survey spectrum of the graphene.

The atomic concentration of oxygen was around 14% and was considered in the peak fit of the C1s spectrum. A comparison between the graphene-on-diamond sample and the diamond substrate is presented in Fig. 1(b). The deconvolution of C1s core level spectrum after background subtraction using Shirley's method is performed on the graphene sample. The results presented here are based on the original data without any shift to the adventitious C1s peak. Carbon sp$^3$ hybridization is on our diamond substrate observed at 284.3 eV and a C1s fit with three Gaussian–Lorenzian mix functions on the graphene-on-diamond sample yields peaks at 283.7, 284.3, 285.1 and 286.6 eV. As the distance of the peak at 283.7 eV is around 0.6 eV from $sp^3$, it is a signature of $sp^2$ carbon bonding [30–32]. The peaks at 285.1 and 286.6 eV are related to oxidized carbon as oxygen was observed in the survey.

The thickness of the catalyst (Ni) is one important parameter that influences the graphene growth. A study on Ni thickness deposited on diamond reports that a narrower G- and 2D-peak corresponding to multilayer graphene is observed with an increased Ni thickness [18]. We investigated Ni thicknesses between 50 and 300 nm with Raman spectroscopy and plotted the ratio between $I_{2D}$ and $I_G$ as probability density functions, as seen in Fig. 2(a). All thicknesses give rise to multilayer graphene but a Ni thickness of 50 nm gives the highest ratio ($I_{2D}/I_G \sim 1.39$) and hence this work focuses only on a Ni thickness of 50 nm. Fig. 2(b) shows an example of monolayer ($I_{2D}/I_G \geq 2$), bilayer ($1 \leq I_{2D}/I_G < 2$) and multilayer ($I_{2D}/I_G < 1$) graphene [19, 28] 2D-Raman mapping was performed and the obtained ratio between $I_{2D}$ and $I_G$ is presented in Fig. 2(c) together with a histogram of the values. The monolayer coverage is estimated to be around 19.8% and the bilayer to approximately 36.2%. It is known that Ni has a high carbon solubility and diffusivity compared to other metal catalysts used for graphene synthesis [33]. The carbon solubility decreases with annealing temperature as reported by Ref. [34]. This could lead to a lower number of graphene layers which in turn cause us to observe a higher monolayer graphene coverage than Ref. [19].

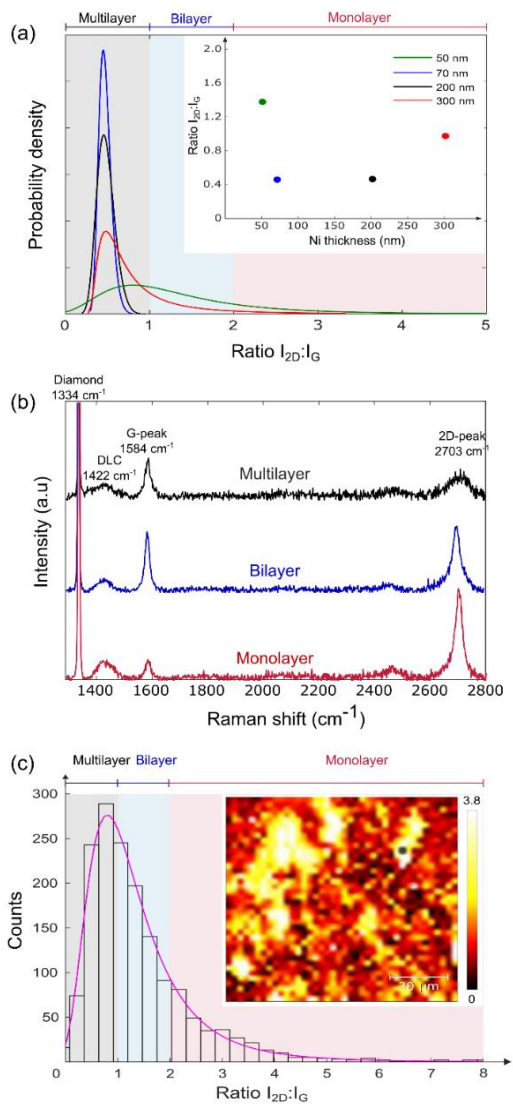

**Fig. 2.** (a) The probability density function of the intensity ratio between $I_{2D}$ and $I_G$ at different Ni thicknesses. The inset shows the average of the probability density of $I_{2D}/I_G$ as a function of Ni thickness. (b) A comparison between monolayer, bilayer and multilayer graphene with a Ni thickness of 50 nm. (c) Histogram of the results from the 2D-Raman mapping of $I_{2D}/I_G$. The inset shows the results of the 2D-Raman mapping of $I_{2D}/I_G$ with a Ni thickness of 50 nm.

In order to perform electrical measurements a two-step standard lithography process was used to create a Hall bar structure as well as 50 nm thick gold contacts on the top of the graphene. Hall-effect measurements were performed at temperatures between 80 and 360 K.

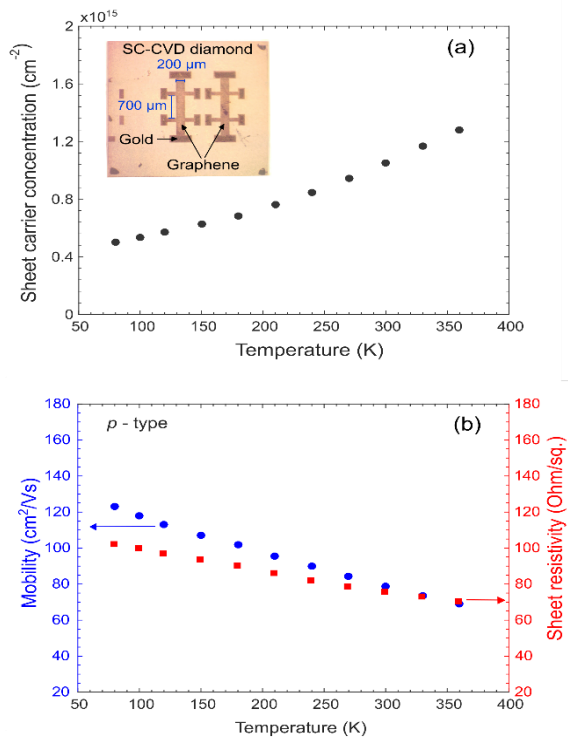

**Fig. 3.** Hall-effect measurements of graphene grown on diamond; plotted across different temperatures. (a) the sheet carrier concentration. The inset shows the Hall-bar configuration. (b) the sheet resistivity (red squares) and the Hall mobility (blue circles).

The Hall mobility measurement reveals that holes are the dominating carriers. Fig. 3 shows that the sheet hole concentration, the sheet resistivity and the Hall mobility at RT were $2.1 \times 10^{13}$ cm$^{-2}$, 3770 $\Omega/\square$ and 79 cm$^2$/Vs, respectively. At 80 K they were instead determined to be $1 \times 10^{13}$ cm$^{-2}$, 5100 $\Omega/\square$ and 123 cm$^2$/Vs, respectively. The Hall mobility is lower than what is given in Ref. [19] (140 cm$^2$/Vs at RT). The ambient conditions at which epitaxial graphene formation is performed have a strong influence on the graphene quality. For example, in previous studies, the Hall mobility of a single graphene layer grown on SiC showed variation from 4 to 150,000 cm$^2$/Vs. The outcome depends on parameters such as the annealing system used (an ultra-high vacuum chamber or a radio frequency furnace), the atmosphere during annealing (vacuum or Ar$_2$), and also the graphene crystalline quality (grain size, carrier concentration, shape, defects etc.) [35–37]. These complex dependencies might also explain the lower mobility observed in our graphene films.

After chemically removing the graphene, an irregular surface roughness across all the samples could be observed (Fig. 4).

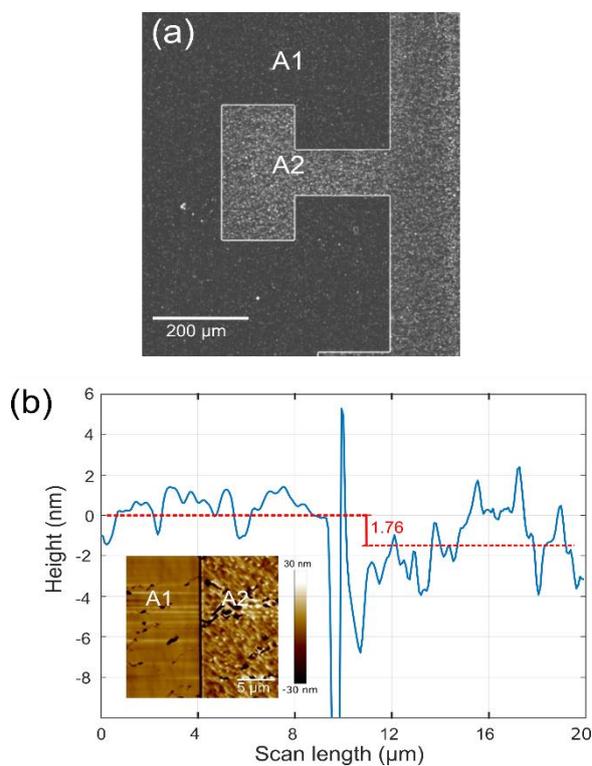

**Fig. 4.** The diamond surface after removing the graphene. (a) Image of the diamond surface obtained from 3D coherence scanning interferometry. The darker color (A1) indicates an area which was not deposited with graphene and the brighter area (A2) had graphene deposited. (b) Averaged AFM results of the step difference between area A1 and A2. The difference in average height between A1 and A2 is around 1.76 nm.

AFM analysis showed that the sample surface before and after the growth process significantly differ. The surface roughness of the area which had not been processed (A1, $9.5 \times 20$ µm) was measured to be below 3 nm while the area where graphene had been deposited (A2, 10 × 20 µm) had a surface roughness varying from 7 to 9 nm. The average step difference over three scanned areas (19.5 × 20 µm) was between 1.76 and 12.3 nm. Fig. 4(b) shows an example of AFM results across the step. The increased surface roughness after deposition can be somewhat attributed to that pits formed on the surface. These likely existed before the deposition (-caused by polishing, dislocations etc.) but became wider and deeper by the process. The increased roughness of the diamond surface probably originates from the carbon-nickel reaction. This phenomenon has previously been observed to occur during annealing in hydrogen and $Ar_2$ atmosphere at 800 to 900 °C [38–41].

The definite relationship between the Ni thickness and the number of graphene layers is not yet established. The thickness of graphene layers strongly depends on the number of carbon atoms diffused through the Ni film. According to Ref. [34], the growth of graphene layers includes at least two processes which can occur only at defined temperature ranges. The first is carbon dissolution in Ni at higher temperatures and the second is graphene precipitation on the Ni surface as the carbon atoms are crystallized. Depending on the Ni thickness one could also differentiate between bulk and surface diffusion. Diffusion barriers are defined by bulk values, where the metal thickness is one of them [42]. However, another study [43] states that carbon atoms can migrate either through the lattice or the grain boundary, and require different activation energies. It is possible that the diffusion paths are through an imperfect crystal lattice. The activation energy depends on the distance between the carbon atom and the metal surface layer. Moreover, the correlation between Ni thickness and the number of graphene layers also strongly depends on the crystallinity of the Ni film, which changes during annealing. The diffusion path is thus also changing. A higher annealing temperature and a longer annealing time increase the carbon solubility and the diffusion length which in turn increases the number of graphene layers as is presented by Ref. [34]. For this reason, the carbon diffusion mechanism and the catalytic reaction between the Ni film and diamond is very complex and require further investigation. Understanding the process will likely enable control of the number of deposited graphene layers.

## Conclusion

We have demonstrated a fast process for deposition of graphene directly on (100) single-crystalline diamond and had this confirmed by Raman spectroscopy and XPS. By using Ni we obtained a mixture of multilayer, bilayer and monolayer graphene where around 20% constituted to monolayer graphene. The coverage distribution was derived from the intensity ratio mapping between the 2D- and G-peak using Raman spectroscopy. The direct growth was performed by a high-temperature (800˚C) annealing process for a duration of 1 min. We were able to produce a millimeter-range large area graphene with a process requiring a combination of a lower temperature and a shorter annealing time than what previous reports indicate [18,19]. A Hall hole mobility around 79 $cm^2$/Vs was calculated at room temperature. The results are very promising to achieve future carbon-carbon heterostructures for many applications.

## Acknowledgments

This study was supported by ÅForsk Foundation (Grant number 19-427 and 21-

53), Lennander Foundation and the Swedish Energy Agency (Grant no: 48591-1). The authors would also like to thank the inorganic chemistry group at Uppsala University for letting us use their Raman spectrometer.